\documentstyle[11pt,epsf]{article}
\makeatletter
\def\revtex@ver{1.6}		
\def\revtex@date{12 Aug 93}	
\def\revtex@org{PASP}		
\def\revtex@jnl{}		
\def\revtex@genre{conference proceedings}	
\def\revtex@pageid{\xdef\@thefnmark{\null}
\@footnotetext{This \revtex@genre\space was prepared with the
\revtex@org\space \revtex@jnl\space Rev\TeX\ macros v\revtex@ver.}}
\ifnum\@ptsize<1
\fi
\def\ps@paspcstitle{\let\@mkboth\@gobbletwo
\def\@oddhead{\null{\footnotesize\it\@slug}\hfil}
\def\@oddfoot{\rm\hfil\thepage\hfil}
\let\@evenhead\@oddhead\let\@evenfoot\@oddfoot
}
\def\ps@myheadings{\let\@mkboth\@gobbletwo
\def\@oddhead{\hbox{}\hfil\sl\rightmark\hskip 1in\rm\thepage}%
\def\@oddfoot{}%
\def\@evenhead{\rm\thepage\hskip 1in\sl\leftmark\hfil\hbox{}}%
\def\@evenfoot{}\def\sectionmark##1{}\def\subsectionmark##1{}}
\def\@leftmark#1#2{\sec@upcase{#1}}
\def\@rightmark#1#2{\sec@upcase{#2}}
\def\@singleleading{0.9}
\def\@doubleleading{1.6}
\def\baselinestretch{\@singleleading}
\def\tightenlines{\def\baselinestretch{\@singleleading}}
\def\loosenlines{\def\baselinestretch{\@doubleleading}}
\clubpenalty\@M
\widowpenalty\@M
\def\@journalname{ASP Conference Series}
\def\cpr@holder{Astronomical Society of the Pacific}
\def\@jourvol{10000}
\def\cpr@year{1994}
\def\vol@title{Astronomical Data Analysis Software and Systems III}
\def\vol@author{R.\ J.\ Hanisch, D.\ R.\ Crabtree, and J.\ Barnes, eds.}
\let\journalid=\@gobbletwo
\let\articleid=\@gobbletwo
\let\received=\@gobble
\let\accepted=\@gobble
\def\@slug{{\tabcolsep\z@\begin{tabular}[t]{l}\vol@title\\
\@journalname, Vol.\ \@jourvol, \cpr@year\\
\vol@author
\end{tabular}}
}
\def\paspconf@frontindent{.45in}
\def\title#1{\vspace*{1.0\baselineskip}
\@tempdima\textwidth \advance\@tempdima by-\paspconf@frontindent
\hfill
\parbox{\@tempdima}
	{\pretolerance=10000\raggedright\large\bf\sec@upcase{#1}}\par
\vspace*{1\baselineskip}\thispagestyle{title}}
\def\author#1{\vspace*{1\baselineskip}
\@tempdima\textwidth \advance\@tempdima by-\paspconf@frontindent
\hfill
\parbox{\@tempdima}
{\pretolerance=10000\raggedright{#1}}\par}
\def\affil#1{\vspace*{.5\baselineskip}
\@tempdima\textwidth \advance\@tempdima by-\paspconf@frontindent
\hfill
\parbox{\@tempdima}
{\pretolerance=10000\raggedright{\it #1}}\par}

\def\abstract{\vspace*{1.3\baselineskip}\bgroup\leftskip\paspconf@frontindent
\noindent{\bf\sec@upcase{Abstract.}}\hskip 1em}
\def\endabstract{\par\egroup\vspace*{1.4\baselineskip}}
\skip\footins 4ex plus 1ex minus .5ex
\long\def\@makefntext#1{\noindent\hbox to\z@{\hss$^{\@thefnmark}$}#1}

\def\tablenotetext#1#2{
\@temptokena={\vspace{.5ex}{\noindent\llap{$^{#1}$}#2}\par}
\@temptokenb=\expandafter{\tblnote@list}
\xdef\tblnote@list{\the\@temptokenb\the\@temptokena}}
\def\spewtablenotes{
\ifx\tblnote@list\@empty
\else
\let\@temptokena=\tblnote@list
\gdef\tblnote@list{\@empty}
\vspace{4.5ex}
\footnoterule
\vspace{.5ex}
{\footnotesize\@temptokena}
\fi}
\newtoks\@temptokenb
\def\tblnote@list{}
\def\endtable{\spewtablenotes\end@float}
\@namedef{endtable*}{\spewtablenotes\end@dblfloat}

\def\thefigure{\@arabic\c@figure}
\def\fnum@figure{Figure \thefigure.}
\def\thetable{\@arabic\c@table}
\def\fnum@table{Table \thetable.}
\long\def\@makecaption#1#2{
\vskip 10pt
\setbox\@tempboxa\hbox{#1\hskip 1.5em #2}
\let\@tempdima=\hsize \advance\@tempdima by -2em
\ifdim \wd\@tempboxa >\@tempdima
	{\leftskip 2em
	#1\hskip 1.5em #2\par}
\else
	\hbox to\hsize{\hskip 2em\box\@tempboxa\hfil}
\fi}
\def\fps@figure{tbp}
\def\fps@table{htbp}
\let\keywords=\@gobble
\let\subjectheadings=\@gobble
\def\upper{\def\sec@upcase##1{\uppercase{##1}}}
\def\sec@upcase#1{\relax#1}
\def\section{\@startsection {section}{1}{\z@}{-4.2ex plus -1ex minus
-.2ex}{2.2ex plus .2ex}{\normalsize\bf}}
\def\subsection{\@startsection{subsection}{2}{\z@}{-2.2ex plus -1ex minus
-.2ex}{1.1ex plus .2ex}{\normalsize\bf}}

\def\subsubsection{\@startsection{subsubsection}{3}{\z@}{-2.2ex plus
-1ex minus -.2ex}{-1.2em}{\normalsize\it}}
\def\thesection{\@arabic\c@section.}
\def\thesubsection{\thesection\@arabic\c@subsection.}
\def\thesubsubsection{\thesubsection\@arabic\c@subsubsection.}
\def\@sect#1#2#3#4#5#6[#7]#8{\ifnum #2>\c@secnumdepth
\def\@svsec{}\else
\refstepcounter{#1}\edef\@svsec{\csname the#1\endcsname\hskip 1em }\fi
\@tempskipa #5\relax
\ifdim \@tempskipa>\z@
\begingroup #6\relax
\@hangfrom{\hskip #3\relax\@svsec}{\interlinepenalty \@M \sec@upcase{#8}\par}%
\endgroup
\csname #1mark\endcsname{#7}\addcontentsline
{toc}{#1}{\ifnum #2>\c@secnumdepth \else
\protect\numberline{\csname the#1\endcsname}\fi
#7}\else
\def\@svsechd{#6\hskip #3\@svsec #8\csname #1mark\endcsname
{#7}\addcontentsline
{toc}{#1}{\ifnum #2>\c@secnumdepth \else
\protect\numberline{\csname the#1\endcsname}\fi
#7}}\fi
\@xsect{#5}}
\def\@ssect#1#2#3#4#5{\@tempskipa #3\relax
\ifdim \@tempskipa>\z@
\begingroup #4\@hangfrom{\hskip #1}{\interlinepenalty \@M \sec@upcase{#5}\par}\endgroup
\else \def\@svsechd{#4\hskip #1\relax #5}\fi
\@xsect{#3}}
\def\acknowledgments{\@startsection{paragraph}{4}{1em}
{1ex plus .5ex minus .5ex}{-1em}{\bf}{\sec@upcase{Acknowledgments.}}}

\def\qanda@heading{Discussion}
\newif\if@firstquestion \@firstquestiontrue

\def\mathwithsecnums{
\@newctr{equation}[section]
\def\theequation{\hbox{\normalsize\arabic{section}-\arabic{equation}}}}
\def\references{\section*{References}
\bgroup\parindent=0pt\parskip=.5ex
\def\refpar{\par\hangindent=3em\hangafter=1}}
\def\endreferences{\refpar\egroup}

\def\@biblabel#1{\relax}
\def\@cite#1#2{#1\if@tempswa , #2\fi}
\def\reference{\relax\refpar}
\def\@citex[#1]#2{\if@filesw\immediate\write\@auxout{\string\citation{#2}}\fi
\def\@citea{}\@cite{\@for\@citeb:=#2\do
{\@citea\def\@citea{,\penalty\@m\ }\@ifundefined
{b@\@citeb}{\@warning
{Citation `\@citeb' on page \thepage \space undefined}}%
{\csname b@\@citeb\endcsname}}}{#1}}
\let\jnl@style=\rm
\def\ref@jnl#1{{\jnl@style#1\/}}
\def\aj{\ref@jnl{AJ}}			
\def\araa{\ref@jnl{ARA\&A}}		
\def\apj{\ref@jnl{ApJ}}			
\def\apjl{\ref@jnl{ApJ}}		
\def\apjs{\ref@jnl{ApJS}}		
\def\ao{\ref@jnl{Appl.Optics}}		
\def\apss{\ref@jnl{Ap\&SS}}		
\def\aap{\ref@jnl{A\&A}}		
\def\aapr{\ref@jnl{A\&A~Rev.}}		
\def\aaps{\ref@jnl{A\&AS}}		
\def\azh{\ref@jnl{AZh}}			
\def\baas{\ref@jnl{BAAS}}		
\def\jrasc{\ref@jnl{JRASC}}		
\def\memras{\ref@jnl{MmRAS}}		
\def\mnras{\ref@jnl{MNRAS}}		
\def\pra{\ref@jnl{Phys.Rev.A}}		
\def\prb{\ref@jnl{Phys.Rev.B}}		
\def\prc{\ref@jnl{Phys.Rev.C}}		
\def\prd{\ref@jnl{Phys.Rev.D}}		
\def\prl{\ref@jnl{Phys.Rev.Lett}}	
\def\pasp{\ref@jnl{PASP}}		
\def\pasj{\ref@jnl{PASJ}}		
\def\qjras{\ref@jnl{QJRAS}}		
\def\skytel{\ref@jnl{S\&T}}		
\def\solphys{\ref@jnl{Solar~Phys.}}	
\def\sovast{\ref@jnl{Soviet~Ast.}}	
\def\ssr{\ref@jnl{Space~Sci.Rev.}}	
\def\zap{\ref@jnl{ZAp}}

\def\deg{\hbox{$^\circ$}}

\def\la{\mathrel{\hbox{\rlap{\hbox{\lower4pt\hbox{$\sim$}}}\hbox{$<$}}}}
\def\ga{\mathrel{\hbox{\rlap{\hbox{\lower4pt\hbox{$\sim$}}}\hbox{$>$}}}}

\newcount\lecurrentfam
\def\LaTeX{\lecurrentfam=\the\fam \leavevmode L\raise.42ex
\hbox{$\fam\lecurrentfam\scriptstyle\kern-.3em A$}\kern-.15em\TeX}

\def\plotfiddle#1#2#3#4#5#6#7{\centering \leavevmode
\vbox to#2{\rule{0pt}{#2}}
\includegraphics{#1}}
\newif\if@finalstyle \@finalstylefalse
\if@finalstyle
\ps@myheadings		
\let\ps@title=\ps@paspcstitle	
\else
\ps@plain			
\let\ps@title=\ps@plain	
\fi
\ds@twoside
\makeatother
\begin{document}

\title{Stellar Orbits in Barred Galaxies with Nuclear Rings}

\author{Clayton H. Heller}
\affil{Universit\"ats Sternwarte, Geismarlandstra\ss e 11, 
       D-37083 G\"ottingen, Germany}
\author{Isaac Shlosman}
\affil{Department of Physics and Astronomy, University of Kentucky,
       Lexington, KY 40506-0055, USA}

\begin{abstract}
We investigate the dynamical response of stellar orbits in a rotating
barred galaxy potential to the perturbation by a nuclear gaseous ring.
The change in 3D periodic orbit families is examined as the gas accumulates
near the inner Lindblad resonance.  
It is found that the $x_2/x_3$ loop extends to higher Jacobi energy
and a vertical instability strip forms in each family.
These strips are connected by a symmetric/anti-symmetric
pair of $2\!\!:\!\!2\!\!:\!\!1$ 3D orbital families. A significant
distortion of the $x_1$ orbits is observed in the vicinity of the ring, which
leads to the intersection between orbits over a large range of the Jacobi
integral. We also find that a moderately elliptical ring oblique to the
stellar bar produces
significant phase shifts in the $x_1$ orbital response.
\end{abstract}

About 2/3 of all disk galaxies are weakly or strongly barred (de Vaucouleurs
1963), many more are ovally distorted (Bosma 1981; Kormendy 1982) or have
triaxial bulges (Kormendy 1994). Central starburst activity in these galaxies 
often
delineates $\sim$few$\times100$ pc size ring-like structures of star forming
regions mixed with molecular gas and dust (Buta \& Crocker 1993). Nuclear
``rings'' seem to be associated with inner Lindblad resonances (ILRs) (Telesco
\& Decher 1988; Shlosman\,{\it et\,al.}\,1989; Kenney\,{\it et\,al.}\,1992; 
Athanassoula 1992; Knapen\,{\it
et\,al.}\,1995a,b). Their intrinsic shapes vary from circular to moderately
elliptical, in which case they lead stellar bars by 
$\sim 50-90\deg.$ As such, nuclear rings are moderately strong
perturbations on the gravitational potential of the central galactic region,
thus affecting stellar orbits and gas flow there.
We analyze the main stellar orbits (in the plane and 3D) in the presence
of a ring (see also Heller \& Shlosman 1995).

The galaxy model consists of the superposition of four components:
disk, bulge, bar, and ring.  The disk is represented by a Miyamoto-Nagai
potential, the bulge by a Plummer sphere, and the bar by a triaxial
Ferrers density distribution.  The ring is centered in the ILR region
and for the models presented here is equivalent to $\sim10^9\,{\rm M}_\odot$
or 38\% of the local mass (Model D).  

The characteristic diagram for the main planar prograde periodic orbits in the
inner region of the model when no ring is present (Model A) is shown in the
upper frame of Figure~1.
\begin{figure}[t]
\plotfiddle{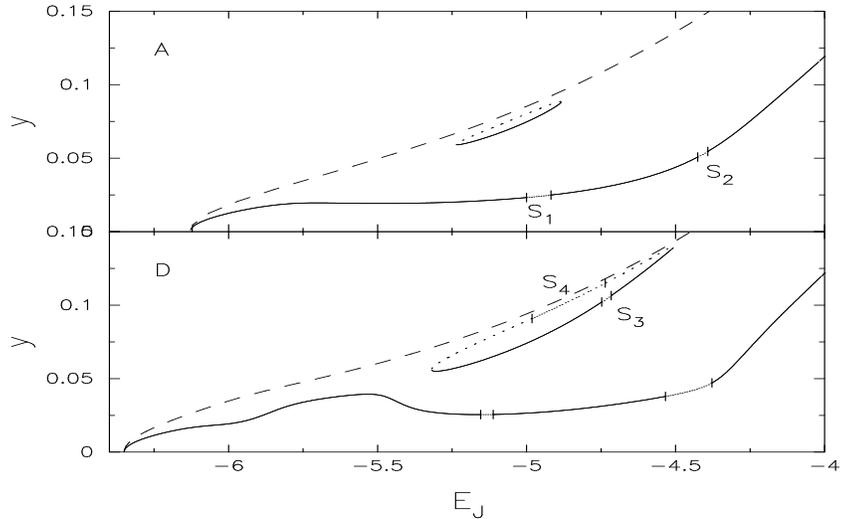}{5cm}{0}{80}{55}{-240}{-200}
\caption{Characteristic diagrams of the $x_1$, $x_2$, and $x_3$ families
for models without (A) and with (D) a nuclear ring.  Stable sections of 
the characteristics are represented by solid lines while unstable are 
broken. Four vertical instability strips are marked. The long-dashed curve
is the zero velocity curve.}
\end{figure}
The three direct families $x_1$, $x_2$, and $x_3$
are shown along with two vertical instability strips.  From the vertical 
instability strips bifurcate pairs of symmetric/anti-symmetric 3D orbital
families, $2\!\!:\!\!2\!\!:\!\!1$ (BAN/ABAN) and $2\!\!:\!\!3\!\!:\!\!1$
families from $S_1$ and $S_2$, respectively.  The $2\!\!:\!\!3\!\!:\!\!1$
families have interesting orbital shapes that are symmetric about one
vertical plane while being anti-symmetric about the corresponding 
perpendicular vertical plane.

As the mass of the ring is increased a ``bump'' in the $x_1$ family
forms and broadens at an ${\rm E}_{\rm J}$ below the ILR.  This distortion
represents a local maximum in the y-extent of the orbits, resulting in
a large region of the $x_1$ family to have orbits that intersect with
other $x_1$ orbits at higher ${\rm E}_{\rm J}$.  Such orbit intersections
also occur in the $x_2$ family as a local maximum in orbit eccentricity
develops along the sequence.
Also, as the ring's mass is increased the region of stability
close to the plane of the $2\!\!:\!\!2\!\!:\!\!1$ symmetric family increases
while the $x_2/x_3$ loop extends to higher ${\rm E}_{\rm J}$ and develops
two regions of vertical instability.  These two instability strips,
one on $x_2$ and one on $x_3$, are connected by a symmetric/anti-symmetric
pair of $2\!\!:\!\!2\!\!:\!\!1$ families elongated perpendicular to the stellar
bar. The symmetric family is stable over half of its characteristic, while
the  anti-symmetric is unstable everywhere.
The bottom frame of Figure~1 shows the planar characteristic diagram 
for the model with a circular ring and indicates the location of the
$x_2/x_3$ instability strips $S_3$ and $S_4$.

In Figure~2a
\begin{figure}[t]
\plotfiddle{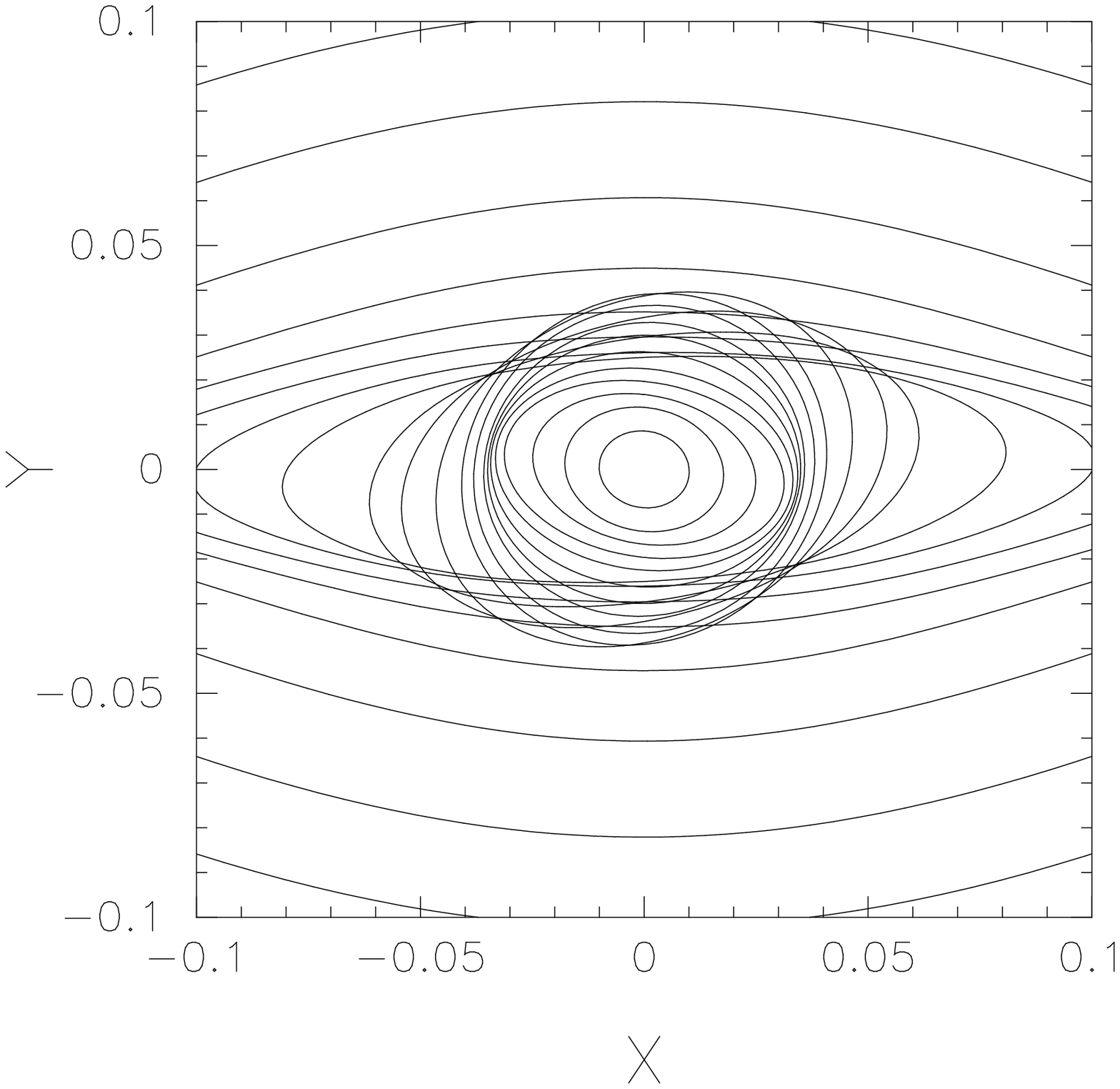}{2cm}{0}{33}{32}{-210}{-115}
\plotfiddle{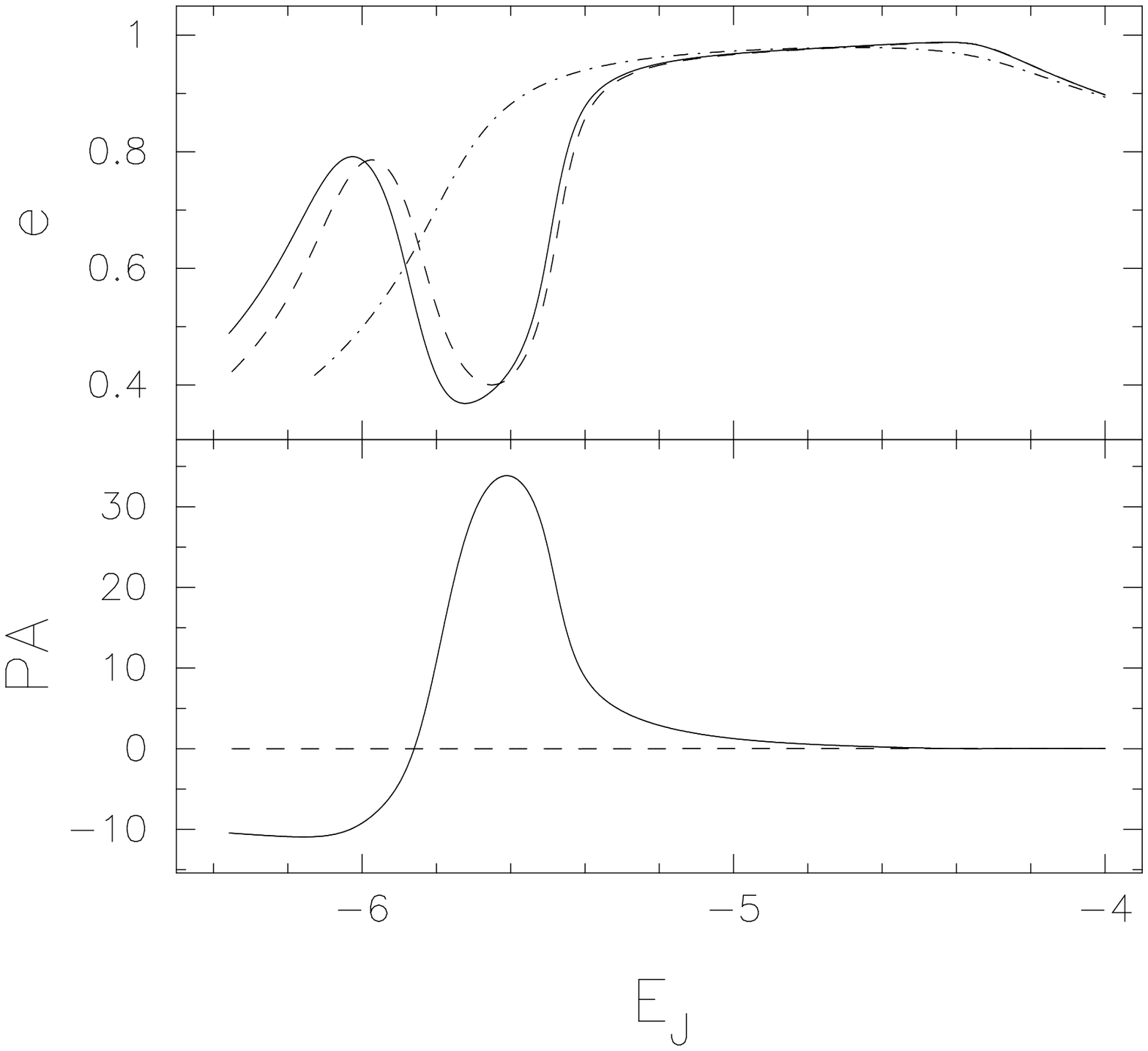}{2cm}{0}{40}{33.1}{-35}{-87}
\caption{(a) Twisting of $x_1$ orbits in model with oblique elliptical
ring. The ring with an ellipticity of 0.4 and semi-major axis
0.04 units is leading the bar by 60\,deg.
The frame is 2\,Kpc on a side.
(b) Eccentricity and position angle of $x_1$ orbits from models 
without ring (dot-dashed), with circular ring (dashed), and oblique
elliptical ring (solid), as a function of the Jacobi energy.}
\end{figure}
we show the phase shift or twisting of the $x_1$ orbital
alignment in response to a moderately elliptical ($e=0.4$) ring
leading the stellar bar by 60 degrees.  The change in ellipticity 
and position angle is given in Figure~2b and is compared 
with models A and D.  It can be seen that while the
eccentricity as a function of ${\rm E}_{\rm J}$ is only slightly offset
from the circular ring case, the position angle of the orbit semi-major 
axis swings from -10 to 35 degrees with respect to the bar. Note, that
innermost $x_1$ orbits trail the bar. 
The interior orbits remain stable and continue
to trap a significant region of phase space around them. 

The main effect of the circular nuclear ring is to produce intersecting orbits
over a wide range of Jacobi energies in both the $x_1$ and $x_2$ 
orbit families.  Gas on such orbits will shock and dissipate energy
on a dynamical time scale.  As a consequence, the gas will quickly 
settle down deep inside the resonance region, further
enhancing the ring. The growth of the ring is limited by its self-gravity.
For a non-circular ring oblique to the stellar
bar and leading it, the twisting of the $x_1$ orbits will further enhance 
shocks in the gas. It is clear from Figure 2, 
that both trailing and leading shocks will develop.

\end{document}